\documentclass[showpacs,aps,pra,twocolumn,amsmath,amssymb,amsfonts,superscriptaddress]{revtex4}
\usepackage{amsmath}
\usepackage[latin1]{inputenc}
\usepackage{graphicx}
\usepackage[dvipdf]{epsfig}
\usepackage{hyperref}

\begin{document}
\title{Equivalence between discrete quantum walk models in
arbitrary topologies}

\author{F. M. Andrade}
\author{M. G. E. da Luz}
\email{luz@fisica.ufpr.br}
\affiliation{Departamento de F\'{\i}sica,
Universidade Federal do Paran\'a,
C.P. 19044, 81531-990 Curitiba-PR, Brazil}

\date{\today}

\begin{abstract}
Coin and scattering are the two major formulations 
for discrete quantum walks models, each believed to have its own 
advantages in different applications.
Although they are related in some cases, it was an open question 
their equivalence in arbitrary topologies.
Here we present a general construction for the two models
for any graph and also for position dependent transition amplitudes.
We then prove constructively their unitary equivalence.
Defining appropriate projector operators, we moreover show
how to obtain the probabilities for one model from the evolution
of the other.
\end{abstract}


\maketitle

\section{Introduction}

In essence, there are three different possible implementations
for quantum walks (QW), all taking place in discrete spaces (graphs).
Two are quantum analogues of Markov chains \cite{crw,STOC01}: 
generalizations of diffusion-like dynamics, where time is continuous 
(CTQW) \cite{pra.1998.58.915}, and discrete time unitary maps 
\cite{crw}, known as coin QW (CQW).
Recently, it has been shown to exist a direct relation between these 
two cases \cite{strauch,childs-arxiv}.
The third (SQW), also a discrete time formulation, is physically    
appealing since it is based on the idea of scattering in multi-port 
interferometers \cite{hillery-2003,hillery-2007}.

Due to the importance of Markov chains in random algorithms, CQW 
and CTQW are extensively used in their quantum versions 
\cite{ambainis-2003}, often being more efficient because the 
exponentially faster hitting times of QW 
\cite{pra.1998.58.915,exp-faster,kempe2}.
Relevant is also the finding that one can implement universal 
quantum computation through scattering processes in a CTQW model 
\cite{childs-prl}.
Regarding applications for SQW, they seem to be particularly 
suitable \cite{searching-s} to solve searching problems 
\cite{searching}.

It is generally believed that CQW, CTQW and SQW have distinct 
advantages in different contexts 
\cite{hillery-2007,kendon-2007, tregenna}.
For instance, universal port gates in quantum computation like 
Hadamard's \cite{nielsen-book} are easier to implemented with CQW, 
the most common formulation in the mathematical and computer 
science literature.
On the other hand, the construction of SQW in graphs of arbitrary
topologies is more direct, and thus conceivably simpler to realize
experimentally.
Furthermore, since it is based on an scattering approach, analytical
techniques are far more developed for SQW than for coin
models \cite{kendon-2007}.

Hence; (i) to formulate in a constructive way both the CQW and 
SQW for graphs of any topology \cite{any-topology} and for position 
dependent quantum amplitudes \cite{position-dependent}; (ii) to 
prove their unitary equivalence in the general case
(previously established only for particular situations 
\cite{hillery-2003,kosik}); and finally (iii) to show how to
obtain the probabilities for one model from the other;
are fundamental because the following.
First, a constructive rather than an abstract formulation makes 
easier concrete implementations of such systems, even for 
arbitrary topologies (e.g., by means of BE condensates 
in optical lattices \cite{manouchehri}).
Second, such results would bridge the gap between two apparent 
distinct quantization schemes for a same class of systems.
Finally, they also would show that the usages for one model are 
equally possible for the other.
The points (i)-(iii) are then the goals of the present paper.

\section{Graph structures and basic definitions}

We assume an undirected simple arbitrary graph \cite{any-topology}, 
whose nodes are labeled in $\mathbb{Z}$.
Its topology is entirely determined by the sets  
$\mathcal{V}_j = \{ j_1, j_2, \ldots, j_{N_j} \}$, which 
represent the $N_j$ nodes connected to the node $j$. 
Thus, if $j_i$ belongs to $\mathcal{V}_j$, then there exists 
exactly one edge between $j$ and $j_i$.
Also, to any node $j$ we associate the set of integers 
$\Lambda_j = \{1, 2, \ldots, N_j\}$.
Each element $\sigma$ of $\Lambda_j$ corresponds to a different 
edge attached to $j$, in a one-to-one relation.
Note that if $j_r$ and $j_s$ have a common edge, then
there are two integers numbers, $\sigma_r$ and $\sigma_s$, 
associated to such edge, one due to $j_r$ and other to $j_s$.

A given mapping (function) on a graph is said locally-adaptable if:
(i) it can be constructed for any graph node $j$;
(ii) for each $j$, it depends only on the edge structure of the
nodes in $\mathcal{V}_j$; and 
(iii) it is always well-defined regardless the number of elements 
in $\Lambda_{\mathcal{V}_j}$.
This is an important concept because if one can establish time
evolution relaying only on locally-adaptable mappings, 
then the resulting dynamics is valid for any graph topology.

So, consider two locally-adaptable functions that direct 
reflect the specific structure of a given graph.
The first, $e: \ \Lambda_j \rightarrow \mathcal{V}_j$, 
associates each $\sigma$ from $\Lambda_j$ to a single
$j_i$ from $\mathcal{V}_j$, such that $e(\sigma; j)$ gives
the node connected to $j$ through the edge labeled $\sigma$ 
with respect to $j$.
For the particular example schematically depicted in Fig. 
1 (a), we have $\mathcal{V}_j = \{ j_1, j_2, j_3, j_4 \}$ and
$\Lambda_j = \{ 1, 2, 3, 4 \}$, thus 
$e(\sigma_1; j) = j_1$, $e(\sigma_2; j) = j_2$, 
$e(\sigma_3; j) = j_3$, and $e(\sigma_4; j) = j_4$,
where each $\sigma_i$ assumes one of the values in $\Lambda_j$.
The second, $\gamma : \ \Lambda_j \rightarrow 
\Lambda_{\mathcal{V}_j}$, maps each $\sigma_i$ from 
$\Lambda_j$ to the single element
$\sigma$ of $\Lambda_{e(\sigma_i; j)}$ representing the 
edge joining $j$ and $e(\sigma_i; j)$.
Again in Fig. 1 (a) we have
$\gamma(\sigma_1; j) = \sigma_k$, 
$\gamma(\sigma_2; j) = \sigma_l$,
$\gamma(\sigma_3; j) = \sigma_m$,
and
$\gamma(\sigma_4; j) = \sigma_n$.
Here $\sigma_k$ belongs to $\Lambda_{j_1} = \{1, 2\}$,
$\sigma_l$ to $\Lambda_{j_2} = \{1, 2, 3\}$, and so forth.
Note that in general
$\gamma(\gamma(\sigma; j); e(\sigma; j)) = \sigma$.

\begin{figure}[top]
\centerline{\psfig{figure=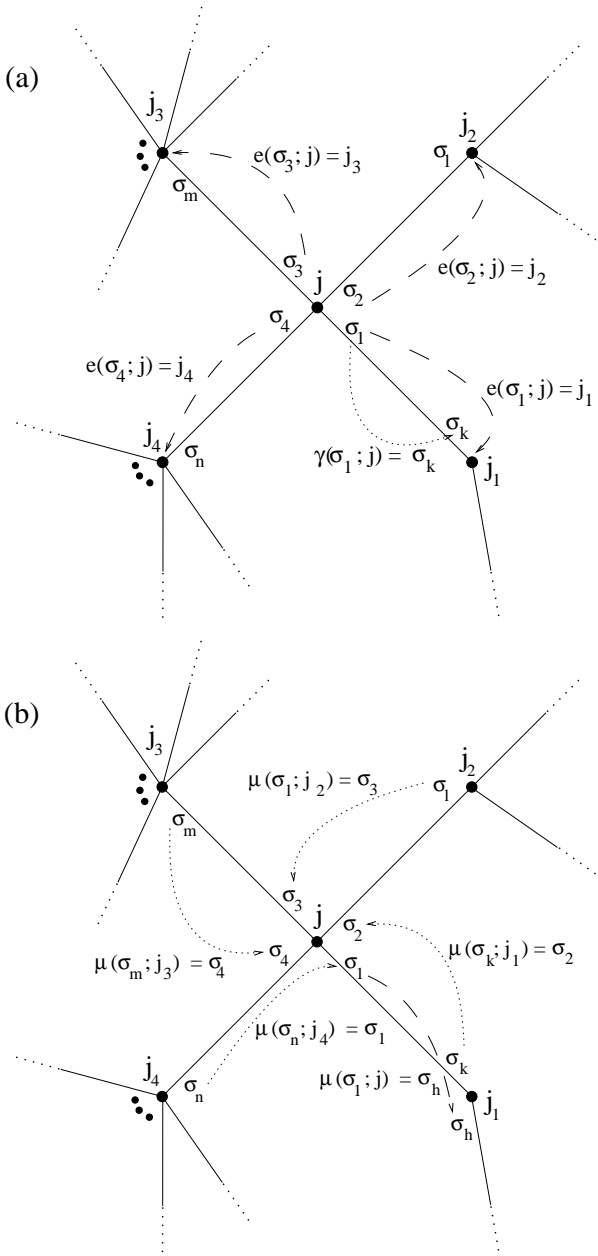,width=8cm}}
\caption{Examples of locally-adaptable mappings.
(a) Mappings which are directly associated to the
graph topology.
(b) Arbitrary mappings, used to define the CQW 
evolution.}
\end{figure}
\medskip

The above mappings are strictly related to the graph specific 
topology. 
However, more general locally-adaptable functions can also 
be defined.
The next three will be very useful.
$\mu : \ \Lambda_j \rightarrow \Lambda_{\mathcal{V}_j}$
extends $\gamma$ since it associates each $\sigma$ from 
$\Lambda_j$ to an unique arbitrary element of 
$\Lambda_{e(\sigma; j)}$, e.g., in the particular
case of Fig. 1 (b), we have
$\mu(\sigma_1; j) = \sigma_h$, where 
$\sigma_h$ is in $\Lambda_{j_1} = \{1, 2\}$.
Although such function is locally-adaptable, for an 
appropriate and consistent latter construction of the quantum 
evolution along the whole graph, we consider an extra 
restriction for $\mu$.
From its very definition, for any $\sigma \in {\Lambda}_j$
we have that $\mu(\gamma(\sigma; j); e(\sigma; j))$ is also an 
element of ${\Lambda}_j$.
Then, we impose additionally that   
$\mu(\gamma(\sigma_r; j); e(\sigma_r; j)) \neq
\mu(\gamma(\sigma_s; j); e(\sigma_s; j))$ 
if $\sigma_r \neq \sigma_s$ 
($\sigma_r, \ \sigma_s$ in ${\Lambda}_j$), i.e.,
the set $\{ \mu(\gamma(\sigma;j); e(\sigma;j)) \} = \Lambda_j$.
Observe that this restriction can always be fullfiled 
whatever the graph topology.
For instance, in Fig. 1 (b) we have
$\{
\mu(\sigma_n ; j_4), \mu(\sigma_k ; j_1),
\mu(\sigma_l ; j_2), \mu(\sigma_m ; j_3) \} =
\{ \sigma_1, \sigma_2, \sigma_3, \sigma_4 \} = {\Lambda}_j$.
Naturally $\mu$ induces two other locally-adaptable 
functions.
Indeed, suppose $\sigma_r$ running over $\Lambda_j$, so
$\Omega_j = \{(\sigma_i, j_i)\} =
\{(\mu(\sigma_r ; j), e(\sigma_r ; j))\}$
($\sigma_i \in \Lambda_{j_i}$ and 
$j_i \in \mathcal{V}_{j}$, ($i = 1, 2, \ldots, N_j$)) is a 
set where to each pair $(\sigma_i, j_i)$ corresponds a distinct 
$\sigma \in \Lambda_{j}$.  
We have thus $\nu : 
\ \Omega_j \rightarrow \Lambda_j$ 
and $a : \ \Omega_j \rightarrow j$, such that
$\nu(\sigma_i; j_i) = \sigma$ and $a(\sigma_i; j_i) = j$.
By construction
\begin{eqnarray}
j &=& 
a(\mu(\sigma; j); e(\sigma; j)) =
e(\nu(\sigma; j); a(\sigma; j)), 
\nonumber \\
\sigma &=& 
\nu(\mu(\sigma; j); e(\sigma; j)) =
\mu(\nu(\sigma; j); a(\sigma; j)). 
\label{mu-nu} 
\end{eqnarray}

\begin{figure}[top]
\centerline{\psfig{figure=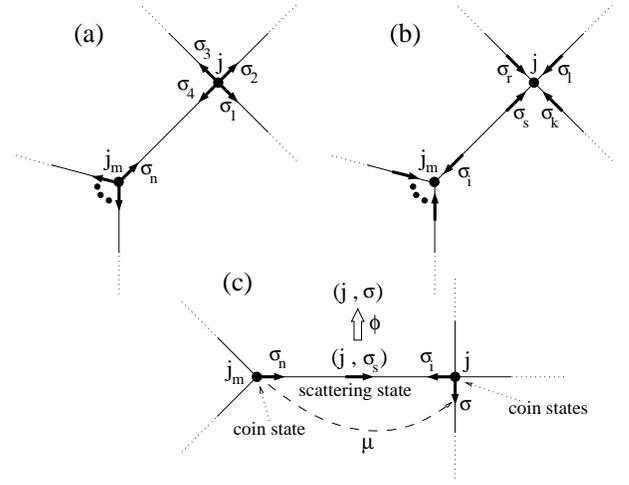,width=8cm}}
\caption{In the coin (a) and scattering (b) QW formulations, 
$\sigma$ is associated, respectively, to nodes and edge states.
(c) An example of relabeling for scattering states,
Eq. (\ref{new-labels}).}
\end{figure}
\medskip

\section{The two discrete time formulations}

For the coin version, the states are defined on the graph 
nodes $j$ \cite{watrous}.
Thus, the $\sigma$'s labeling the edges attached to 
$j$ can be associated to the quantum numbers representing
the different ``outgoing'' directions leaving $j$, as 
schematically shown in Fig. 2 (a).
Hence, we have as the base states $\{ |j, \sigma \rangle_c \}$, 
where for each $j$, $\sigma = 1, 2, \ldots, N_j$ and
$\langle \sigma', j' | j'', \sigma'' \rangle_{c} =
\delta_{j' \, j''} \, \delta_{\sigma' \, \sigma''}$, which 
spans the Hilbert space 
$\mathcal{H} = L^2(\mathbb{Z} \times \mathbb{Z}_{N_{\mathbb{Z}}})$.

To establish the system dynamics, we first consider the shift 
operator $S$ (and its adjoint $S^{\dagger}$ 
\cite{note-adjoint}), such that
\begin{eqnarray}
S |j, \sigma \rangle_c &=& 
|e(\sigma; j), \mu(\sigma; j) \rangle_{c}
\nonumber \\
S^{\dagger} |j, \sigma \rangle_c &=& 
|a(\sigma; j), \nu(\sigma; j) \rangle_{c}.
\end{eqnarray}
From Eq. (\ref{mu-nu}), it follows that 
$S^{\dagger} \, S = S\, S^{\dagger} = {\mathbb I}$ in $\mathcal{H}$.
Then, for each $j$, let $C^{(j)}$ to be a ``coin'' 
operator, represented by a $N_j \times N_j$ unitary matrix, whose 
action over a basis state (of quantum number $j$) is
$C^{(j)} |j, \sigma \rangle_{c} = \sum_{\sigma' = 1}^{N_j}
c^{(j)}_{\sigma' \, \sigma} |j, \sigma' \rangle_{c}$.
Finally, we set the unitary one step time evolution as
\cite{theta}
\begin{equation}
U_c  = S \sum_{j} \sum_{\sigma = 1}^{N_j}
C^{(j)} {|j, \sigma \rangle} \langle \sigma, j |_c, 
\label{uc}
\end{equation}
which is valid for any topology (encoded in the functions 
$e$ and $\gamma$) and defines a very general time evolution for 
the problem through the functions $\mu$, $\nu$ and $a$.

For the scattering version, 
note first that even for a same graph, the $\sigma$ 
labeling \cite{labeling} for the scattering  
\cite{hillery-2003,hillery-2007} can be completely 
distinct than that for the coin formulation 
(see Figs. 2 (a) and (b)). 
Thus, in principle the functions $e$ and $\gamma$ can 
assume different values in the two cases (and they will be 
distinguished when necessary).

Now, two quantum states are defined along each edge, e.g.,
for the edge connecting the nodes $j$ and 
$j_m$ in Fig. 2 (b), we denote the state ``incoming'' to the 
node $j$ ($j_m$) by $| j, \sigma_s \rangle_s$ 
($| j_m, \sigma_i \rangle_s$), which can be written also as
$| e(\sigma_i; j_m), \gamma(\sigma_i; j_m) \rangle_s$
($| e(\sigma_s; j), \gamma(\sigma_s; j) \rangle_s$).
Actually, for any edge, if one state is   
$| j, \sigma \rangle_s$, the other is given by
$| e(\sigma; j), \gamma(\sigma; j) \rangle_s$.
So, the basis set is $\{ |j, \sigma \rangle \}_s$, 
spanning the Hilbert space 
$\mathcal{H} = L^2(\mathbb{Z} \times \mathbb{Z}_{N_{\mathbb{Z}}})$,
as in the coin case.

For the dynamics, we set $U_s = R + T$ \cite{theta}, where the action
of the operators $R$ and $T$ are given by \cite{note-adjoint}
\begin{eqnarray}
R |j, \sigma \rangle_s &=&
r_{\sigma,  \, \sigma}^{(j)}  \,
| e(\sigma; j), \gamma(\sigma; j) \rangle_s,
\nonumber \\
T |j, \sigma \rangle_s &=& 
\sum_{\alpha \in \Lambda_{j}, \, \alpha \neq \sigma} 
t_{\alpha,  \, \sigma}^{(j)}  \,  
| e(\alpha; j), \gamma(\alpha; j) \rangle_s,
\nonumber \\
R^{\dagger} |j, \sigma \rangle_s &=&
r_{\gamma(\sigma; j),  \, \gamma(\sigma; j)}^{(e(\sigma; j)) \, \mbox{*}}  \,
| e(\sigma; j), \gamma(\sigma; j) \rangle_s,
\nonumber \\
T^{\dagger} |j, \sigma \rangle_s &=&
\sum_{\alpha \in \Lambda_{e(\sigma; j)}, \, \alpha \neq \gamma(\sigma; j)}
t_{\gamma(\sigma; j),  \, \alpha}^{(e(\sigma; j)) \, \mbox{*}}  \,  
| e(\sigma; j), \alpha \rangle_s.
\end{eqnarray}
We also define $N_j \times N_j$ scattering matrices 
$\Gamma^{(j)}$, such that
$\Gamma^{(j)}_{\sigma \, \sigma} = r_{\sigma, \sigma}^{(j)}$
and $\Gamma^{(j)}_{\sigma' \, \sigma} = 
t_{\sigma', \sigma}^{(j)}$ (for both $\sigma' \neq \sigma$ in $\Lambda_j$).
If for all $j$ we impose that $\Gamma^{(j)}$ is unitary, then
the coefficients $r$ and $t$ satisfy the usual relations in 
scattering theory \cite{chadam,schmidt}.
So, $U_s$ is unitary.

\section{Obtaining the probabilities}

For QW, stochasticity -- in the classical sense -- comes into play 
only through measurements, when one calculates the probabilities 
for the walker to be found in different locations along the 
graph \cite{kempe1}.
Suppose we shall know at time $n$ what is the probability 
$p_j(n)$ to be in the position state $j$ (which means a node (edge) 
in the coin (scattering) model), regardless the value of the 
coin (direction) quantum number $\sigma$.
So, we define the scattering and coin projector operators as
\begin{eqnarray}
P^{(j, \sigma)}_{\mbox{\scriptsize s}} &=& 
| j, \sigma \rangle \langle \sigma, j |_s +
| e(\sigma; j), \gamma(\sigma; j) \rangle 
\langle \gamma(\sigma; j), e(\sigma; j) |_s,
\nonumber \\
P^{(j)}_{\mbox{\scriptsize c}} &=& 
\sum_{\sigma = 1}^{N_j}
| j, \sigma \rangle \langle \sigma, j |_c. 
\label{projectors}
\end{eqnarray}
The desired probability is thus the expected value 
\begin{equation}
p^{(j)}(n) = \langle \Psi(n) | P | \Psi(n) \rangle,
\ \ \ | \Psi(n) \rangle = U^n | \Psi(0) \rangle 
\label{projection}
\end{equation}
for $P$ one of the expressions in Eq. (\ref{projectors}).

\section{Proving the equivalence of the two formulations}

For so, three steps are necessary: 
(a)  to establish a correspondence between the different
walks states; 
(b) to properly associate their time evolutions;
and 
(c) to construct projector operators to obtain the probabilities
of one in terms of the other.  

Regarding (a), note that we always can define a
locally-adaptable function $\varphi: 
\Lambda_j \rightarrow \Lambda_j$, which for each node $j$, maps 
the quantum number $\sigma$ associated to a specific edge in
the scattering formulation to the quantum number $\sigma'$ 
labeling the same edge, but in the coin formulation.
For example, for the situation in Fig. 2, we have
$\varphi(\sigma_k; j) = \sigma_1$, 
$\varphi(\sigma_l; j) = \sigma_2$,
$\varphi(\sigma_r; j) = \sigma_3$, and
$\varphi(\sigma_s; j) = \sigma_4$.
Also, the actual $\sigma$'s values are not relevant.
They are just a way to label nodes and edges states.
Thus, without loss of generality, for any $j$ we always
can rename one of the model states by 
$|j, \sigma \rangle \Rightarrow |j, \phi(\sigma; j) \rangle $ 
for $\phi$ a bijection $\Lambda_j \rightarrow \Lambda_j$. 
Choosing to retag the scattering case, we consider the
following particular $\phi$
\begin{equation} 
\phi(\sigma; j) = 
\mu(\gamma_c(\varphi(\sigma; j); j);e_c(\varphi(\sigma; j); j)), 
\label{new-labels}
\end{equation}
whose "action" is pictorially represented in Fig. 2 (c).

Then, using this new notation for the scattering states, one has the 
isomorphic unitary operator 
$E: {\mathcal H} \rightarrow {\mathcal H}$~\cite{hillery-2003}
\begin{equation}
E \, |j, \sigma \rangle_s = 
|j, \sigma \rangle_c.
\label{states-mapping}
\end{equation}
associating the scattering state $\sigma$ incoming to $j$ to 
the coin state $\sigma$ outgoing from $j$ (see Fig. 2 (c)).

For point (b), from the explicit form of $U_c$ and
$U_s$, one finds that by setting one of the two 
akin relations
\begin{eqnarray}
\Gamma^{(j)}_{\sigma_b \, \sigma_a} &=&  
c^{(j)}_{\gamma_c(\nu(\sigma_b; j); a(\sigma_b; j)) \, \sigma_a},
\nonumber \\
c^{(j)}_{\sigma_b \, \sigma_a} &=&
\Gamma^{(j)}_{\mu(\gamma_c(\sigma_b; j); e_c(\sigma_b; j)) 
\, \sigma_a}, 
\label{correspondence}
\end{eqnarray}
then, in both models the time evolution transition probability 
amplitudes are exactly the same.
In this case, the resulting dynamics are unitary equivalent since
\begin{equation}
U_{\mbox{\scriptsize{s}}} = E^{\dagger} \, U_{\mbox{\scriptsize{c}}} \, E.
\label{unitary-equivalence}
\end{equation}

Finally, for (c) even if the two formulations are unitarily 
related, the resulting probabilities -- through projections -- 
are not \cite{hillery-2003}.
This is so because each description assumes distinct spatial 
configurations, nodes or edges, to characterize the system,
thus not leading to same probabilities.
In fact, in each edge the two scattering states are mapped by 
$E$ to coin states in different nodes.
Thus, the probability, Eq. (\ref{projection}), to be in a 
unique node is not equal to the probability to be in a unique 
edge.

However, there is a very direct way to obtain the walk 
probabilities for the coin (scattering) model from the scattering 
(coin) model.
We just define
\begin{equation}
P^{(j)}_{\mbox{\scriptsize s}}\big|_{\mbox{\scriptsize c}} = 
E^{\dagger} \, P^{(j)}_{\mbox{\scriptsize c}} \, E, 
\qquad
P^{(j)}_{\mbox{\scriptsize c}}\big|_{\mbox{\scriptsize s}} =
E \, P^{(j)}_{\mbox{\scriptsize s}} \, {E}^{\dagger}.
\label{projections-mapping}
\end{equation}
Then, suppose we construct a SQW with arbitrary $r$'s and $t$'s.
By using $P_s^{(j, \sigma)}$ of Eq. (\ref{projectors})
into Eq. (\ref{projection}), we get the scattering walker 
probabilities at the step $n$.
But now if we use  
$P^{(j)}_{\mbox{\scriptsize s}}\big|_{\mbox{\scriptsize c}}$
for this system, the resulting probabilities are exactly those 
from a coin model, for which the coin matrices elements 
are given by such $r$'s and $t$'s values according to the
correspondence in Eqs. (\ref{correspondence}).
The other way around, to get the scattering model results from 
the CQW, follows in the same fashion.

\section{Remarks and Conclusion}

Although there are some few discussions in the literature on 
how to formulate discrete random walks in general terms 
\cite{any-topology,kendon-2007,kosik}, here we have 
developed a explicit procedure that allows one to write, 
in a constructive way, the time evolution operator directly 
from the system topology and local dynamics.
It also includes the case of position dependent quantum 
amplitudes (through the $C^{(j)}$'s and $\Gamma^{(j)}$'s).
Moreover, in the CQW case, it is not necessary all
the matrices $C^{(j)}$ to have the dimensions equal to 
the largest coordination number of the graph, as in certain 
formulations \cite{kendon-2007}.

In the present framework, a regular graph can be defined in 
terms of the $\Lambda_j$'s sizes  
(e.g., all equal to $N$) and the features of 
$e$, $\gamma$ and $\mu$ (e.g., independent on the $j$'s 
and having specific patterns along the graph).
For instance, for all nodes with the same number of edges,
we know that for the coin, $\mathcal{H}$ can be written
as the direct product of two subspaces, thus
$|j, \sigma \rangle \rightarrow  | \sigma \rangle \otimes 
|j \rangle$.
So, we naturally find from Eq. (\ref{uc}) that
$\sum_j \sum_{\sigma = 1}^{N_j}
C^{(j)} {|j, \sigma \rangle} \langle \sigma, j |
= \sum_j C^{(j)} \otimes |j \rangle \langle j |$,
as in \cite{linden}. 

Usually, it is believed that analytical methods are easier to
implement for the SQW than for CQW \cite{kendon-2007, hillery-analytical}.
Since in fact they can be mapped each other, the existing 
methods for the former should be extendable to the latter.
Our constructive approach may serve as a guide to implement
such extensions \cite{andrade}.

Lastly, by using the cross operators in Eq. 
(\ref{projections-mapping}), we have been able to calculate 
-- for different examples -- the probabilities for SQW and CQW 
from a single implementation.
This will be communicated elsewhere \cite{andrade}.

\section{acknowledgement}

Research grants are provided by Capes and CNPq.


\begin{thebibliography}{mt1}
\bibitem{note-adjoint}
In agreement with the general definition of the adjoint 
$O^{\dagger}$ of a  bounded linear operator $O$:
for any $\varphi$ and $\psi$ in $\mathcal{H}$, 
it holds that $(\varphi, O \psi) = (O^{\dagger} \varphi, \psi)$.
\bibitem{labeling}
Without loss of generality we assume a same $j$ labeling.
\end{thebibliography}
\end{document}